%% file: main.tex
\documentclass[%
 reprint,
 amsmath,amssymb,
 aps,
]{revtex4-1}

\usepackage{graphicx}% Include figure files
\usepackage{dcolumn}% Align table columns on decimal point
\usepackage{bm}% bold math
\usepackage[utf8]{inputenc}
\usepackage[T1]{fontenc}
\usepackage{mathptmx}
\usepackage{import}
\usepackage{xcolor}

\usepackage[acronyms, nonumberlist, nogroupskip,toc]{glossaries}
\newacronym[longplural={Avalanche PhotoDiodes}]{apd}{APD}{Avalanche PhotoDiode}
\newacronym[longplural={Beam Splitters}]{bs}{BS}{Beam Splitter}
\newacronym[longplural={Coincidence-to-Accidentals Ratios}]{car}{CAR}{Coincidence-to-Accidentals Ratio}
\newacronym[longplural={PhotoDiodes}]{pd}{PD}{PhotoDiode}
\newacronym[longplural={Single-Mode Fibers}]{smf}{SMF}{Single-Mode Fiber}
\newacronym{ar}{AR}{Anti-Reflection}
\newacronym{adc}{ADC}{Analog-to-Digital Converter}
\newacronym{bg}{bg}{background}
\newacronym{cw}{CW}{Continuous-Wave}
\newacronym{dcm}{DCM}{4-(dicyanomethylene)-2-methyl-6-(p-dimethylaminostyryl)-4H-pyran}
\newacronym{dc}{DC}{Direct Current}
\newacronym{dm}{DM}{Dichroic Mirror}
\newacronym{dmd}{DMD}{Digital Micromirror Device}
\newacronym{dna}{DNA}{DeoxyriboNucleic Acid}
\newacronym{egfp}{EGFP}{Enhanced Green Fluorescent Protein}
\newacronym{em}{EM}{ElectroMagnetic}
\newacronym{emccd}{EMCCD}{Electron-Multiplying CCD}
\newacronym{fm}{FM}{Flip-Mirror}
\newacronym{fvb}{FVB}{Full Vertical Binning}
\newacronym{fwhm}{FWHM}{Full Width at Half Maximum}
\newacronym{fwm}{FWM}{Four-Wave Mixing}
\newacronym{gbp}{GBP}{Gain-Bandwidth Product}
\newacronym[longplural={Horizontal Shift Speeds}]{hss}{HSS}{Horizontal Shift Speed}
\newacronym{hwp}{HWP}{Half-Wave Plate}
\newacronym{lbo}{LBO}{Lithium triBOrate}
\newacronym{ln2}{LN2}{Liquid Nitrogen}
\newacronym{lp}{LP}{Linearly Polarized}
\newacronym{mfd}{MFD}{Mode-Field Diameter}
\newacronym{mmf}{MMF}{Multi-Mode Fiber}
\newacronym{na}{NA}{Numerical Aperture}
\newacronym{nd}{ND}{Neutral-Density}
\newacronym{nist}{NIST}{National Institute of Standards and Technology}
\newacronym{ne}{NE}{Noise-Eater}
\newacronym{npl}{NPL}{National Physical Laboratory}
\newacronym{nrf}{NRF}{Noise-Reduction Factor}
\newacronym{oa}{OA}{Operational Amplifier}
\newacronym[longplural={Optical Densities}]{od}{OD}{Optical Density}
\newacronym{pbs}{PBS}{Polarizing Beam Splitter}
\newacronym[longplural=Printed Circuit Boards]{pcb}{PCB}{Printed Circuit Board}
\newacronym{pcf}{PCF}{Photonic Crystal Fiber}
\newacronym{pc}{PC}{PhotoConductive}
\newacronym{ppktp}{PPKTP}{Periodically-Poled Potassium-Titanyl-Phosphate}
\newacronym{psd}{PSD}{Power Spectral Density}
\newacronym{pv}{PV}{PhotoVoltaic}
\newacronym{sfwm}{SFWM}{Stimulated Four-Wave Mixing}
\newacronym{shg}{SHG}{Second-Harmonic Generation}
\newacronym{slm}{SLM}{Spatial Light Modulator}
\newacronym{snl}{SNL}{Shot-Noise Limit}
\newacronym{snr}{SNR}{Signal-to-Noise Ratio}
\newacronym[longplural={Silicon Photodiode Detectors}]{spd}{SiPD}{Silicon Photodiode Detector}
\newacronym{spdc}{SPDC}{Spontaneous Parametric DownConversion}
\newacronym{spfwm}{SpFWM}{Spontaneous Four-Wave Mixing}
\newacronym{spm}{SPM}{Self-Phase Modulation}
\newacronym{sprs}{SpRS}{Spontaneous Raman Scattering}
\newacronym{srm}{SRM}{Standard Reference Method}
\newacronym{srs}{SRS}{Stimulated Raman Scattering}
\newacronym{sted}{STED}{STimulated Emission Depletion}
\newacronym{tpe}{TPE}{Two-Photon Excitation}
\newacronym{uv}{UV}{UltraViolet}
\newacronym[longplural={Vertical Shift Speeds}]{vss}{VSS}{Vertical Shift Speed}
\newacronym{wdm}{WDM}{Wavelength-Division Multiplexing}
\newacronym{xpm}{XPM}{Cross-Phase Modulation}
\newacronym{zdw}{ZDW}{Zero-Dispersion Wavelength}

\newcommand{\var}[1]{\textrm{Var}\left[{#1}\right]}

\newcommand{\mean}[1]{\textrm{E}\left[{#1}\right]}
\newcommand{\mse}[1]{\textrm{MSE}\left[{#1}\right]}
\newcommand{\cov}[2]{\textrm{Cov}\left[#1,#2\right]}

\newenvironment{minipeqn}[1][]{\begin{minipage}[#1]{.486\columnwidth}\begin{equation}}{\end{equation}\end{minipage}}

\begin{document}

\title{A practical model of twin-beam experiments for sub-shot-noise absorption measurements}

\author{Jason D. Mueller}
 \email{jason.mueller@bristol.ac.uk}
\author{Nigam Samantaray}
\author{Jonathan C. F. Matthews}
 \email{jonathan.matthews@bristol.ac.uk}
\affiliation{Quantum Engineering Technology Labs, H. H. Wills Physics Laboratory and Department of Electrical and Electronic Engineering, University of Bristol, BS8 1FD, United Kingdom}

\date{\today}

\begin{abstract}
Quantum-intensity-correlated twin beams of light can be used to measure absorption with precision beyond the classical shot-noise limit. The degree to which this can be achieved with a given estimator is defined by the quality of the twin-beam intensity correlations, which is quantified by the noise reduction factor. We derive an analytical model of twin-beam experiments, incorporating experimental parameters such as the relative detection efficiency of the beams, uncorrelated optical noise, and uncorrelated detector noise. We show that for twin beams without excessive noise, measured correlations can be improved by increasing the detection efficiency of each beam, notwithstanding this may unbalance detection efficiency. However, for beams with excess intensity or other experimental noise, one should balance detection efficiency, even at the cost of reducing detection efficiency -- we specifically define these noise conditions and verify our results with statistical simulation. This has application in design and optimization of absorption spectroscopy and imaging experiments.
\end{abstract}

\maketitle

\section{\label{sec:intro}Introduction}

Optical shot-noise is present in all classical imaging and spectroscopy applications using light, and can limit the measurement precision of a parameter once all other technical noise sources have been accounted for~\cite{Celebrano2011,Kukura2010,Chien2018,Miyazaki2014,Ozeki2010,Betzig1986}. Using quantum-intensity-correlated light beams (\textit{i.e.} twin beams)~\cite{Fabre1987(1),Fabre1987(2),Moreau2017,Brida2010,Losero2018} is one method to surpass this classical limit and obtain greater absorption-measurement precision for a given optical power~\cite{Jakeman1986}. Experiments demonstrating this concept have been performed at near-infrared wavelengths using approximately wavelength-degenerate twin beams from downconversion~\cite{Moreau2017,Whittaker2017,Samantaray2017,Chesterking2017,Losero2018,Brida2010}. Monochromatically-pumped \gls{fwm} generates energy-conserving twin beams that are of non-degenerate wavelengths that straddle the pump wavelength. This is useful for imaging and spectroscopy applications because \gls{fwm} can be implemented with a range of materials and pump wavelengths, providing access to a range of twin-beam wavelengths above and below the near-infrared~\cite{Chen2013,Sevigny2015,Pourbeyram2015,Pourbeyram2016,Kowligy2018,Sebbag2019}. However, measurement of highly non-degenerate correlated beams can result in unbalanced detection efficiency, with uncorrelated optical and detector noise present regardless of wavelength degeneracy.

Previous work on measuring quantum intensity correlations from wavelength-degenerate spontaneous downconversion sources have approximated that (1) because the wavelengths are degenerate, so is the loss and detection efficiency of both beams, and (2) there is negligible excess optical or detector noise~\cite{Moreau2017,Brida2010,Vasilyev2000,Bondani2007,Samantaray2017}. Under these assumptions, measured intensity correlations scale with channel efficiency as $1-\eta$, and can always be improved by reducing loss or improving detector efficiency. Here, we show that if either assumption is not true, the scaling of measured intensity correlations depends on the relative twin-beam detection efficiency and properties of the excess noise, and correlations may be improved by reducing the efficiency of one detection channel or unbalancing detection.

In this paper, we present a general analytical framework for twin beam experiments characterized by intensity-difference measurements, extending previous work based on detector calibration~\cite{Berchera2010} and high-power twin beams~\cite{Iskhakov2016}. Our model outputs the \gls{nrf}, a quantifier of twin-beam correlations~\cite{Losero2018,Brida2010,Moreau2017,Finger2015,Berchera2010,Iskhakov2016}, with unbalanced detection loss, uncorrelated optical noise, and uncorrelated detector noise as variables. It is agnostic to the sources of the uncorrelated noise and only requires basic experimental characterization of their mean intensity and variance. We evidence that to improve the quality of measure twin-beam correlations, one should either maximize detection efficiency of both beams, or balance detection efficiency, depending on the properties of the experimental noise. We then confirm our model with statistical simulations, and give a specific real-world example modeling a \gls{fwm} experiment.

\section{\label{sec:nrfTheory}Analytic model of twin-beam intensity correlations including optical and detector noise}

The Fano factor~\cite{Fano1947,RalphBachor} quantifies the intensity noise of a single optical beam, labeled $i$, according to
\begin{equation}
    F_i=\frac{\var{N_i}}{\mean{N_i}},
\end{equation}
where $N_i$ is the random variable associated with the beam's photon number (\textit{i.e.} intensity), characterized by variance $\var{N_i}$ and mean value $\mean{N_i}$. Classically-accessible super-Poissonian intensity fluctuations correspond to $F_i>1$, while Poisson-distributed statistics, which can be achieved by measuring the intensity of a coherent state, correspond to the classical limit of $F_i=1$. Individual beams exhibiting $0\leq F_i<1$ are uniquely non-classical and classified as sub-Poissonian.

The Fano factor can be an important quantifier when searching for optical beams for parameter estimation, as the intensity noise of a probe beam maps onto the uncertainty of estimating a physical parameter, such as absorption~\cite{Whittaker2017}. A beam with $F<1$ is a resource for measuring absorption with precision beyond the classical limit. It is also possible to use a beam with $F\geq1$ for measuring parameters with precision beyond the classical limit, provided the beam is sufficiently well-correlated to another beam that can be measured~\cite{Brida2010,Moreau2017}. 

To quantify the mutual noise characteristics of two beams, we use the \gls{nrf}, given as~\cite{Losero2018,Brida2010,Moreau2017,Finger2015,Berchera2010,Iskhakov2016}
\begin{equation}\label{eqn:noiseReductionFactor}
    \sigma=\frac{\var{N_1-N_2}}{\mean{N_1+N_2}}.
\end{equation}
Note that some authors use a modified form of the \gls{nrf}, where $N_2\rightarrow \left(\mean{N_1}/\mean{N_2}\right)N_2$ to account for detection-efficiency mismatch~\cite{Losero2018,Iskhakov2016}. Our analysis assumes the form in Eq.~\ref{eqn:noiseReductionFactor}, and may be straightforwardly modified to accommodate this alternate \gls{nrf} definition. 

Values of $\sigma\geq1$ correspond to separable classical beams, with $\sigma=1$ representing the classical limit of two Poisson-distributed beams. Values of $0\leq\sigma<1$ correspond to non-classical twin-beam intensity correlations. However, this alone is insufficient to achieve parameter estimation with precision better than what can be achieved with a single pass of a single beam with $F=1$, defined as the classical \gls{snl}~\cite{Brida2010,Moreau2017,Losero2018}. Sub-\gls{snl} parameter estimation is also linked to choice of estimator (see Appendix A).

We describe two correlated twin beams $i=1,2$ with intensity mean and variance
\begin{gather}
    \mean{N_i}=\eta_i\mean{N}\label{eqn:beamMean}\\[6pt]
    \var{N_i}=\mean{N_i}+\beta\mean{N_i}^2\label{eqn:superPoissonBeamVar},
\end{gather}
where $N$ is the lossless photon number, equal among both beams due to the energy-conserving nature of twin-beam production. The parameter $0\leq\eta_i\leq1$ is the total efficiency of each beam's optical path (\textit{i.e.} channel efficiency), comprising loss from all optical components and detection efficiency. The parameter $\beta\geq0$ is used to account for super-Poissonian intensity fluctuations of the individual beams of the twin-beam system~\cite{Purcell1956,Mandel1958,Mandel1959,Mandel1986}, and is equivalent to the second-order intensity correlation function $g^{(2)}(0)-1$~\cite{Mandel1986}.

\begin{figure}[h]
	\centering
	\def\svgwidth{\columnwidth}
	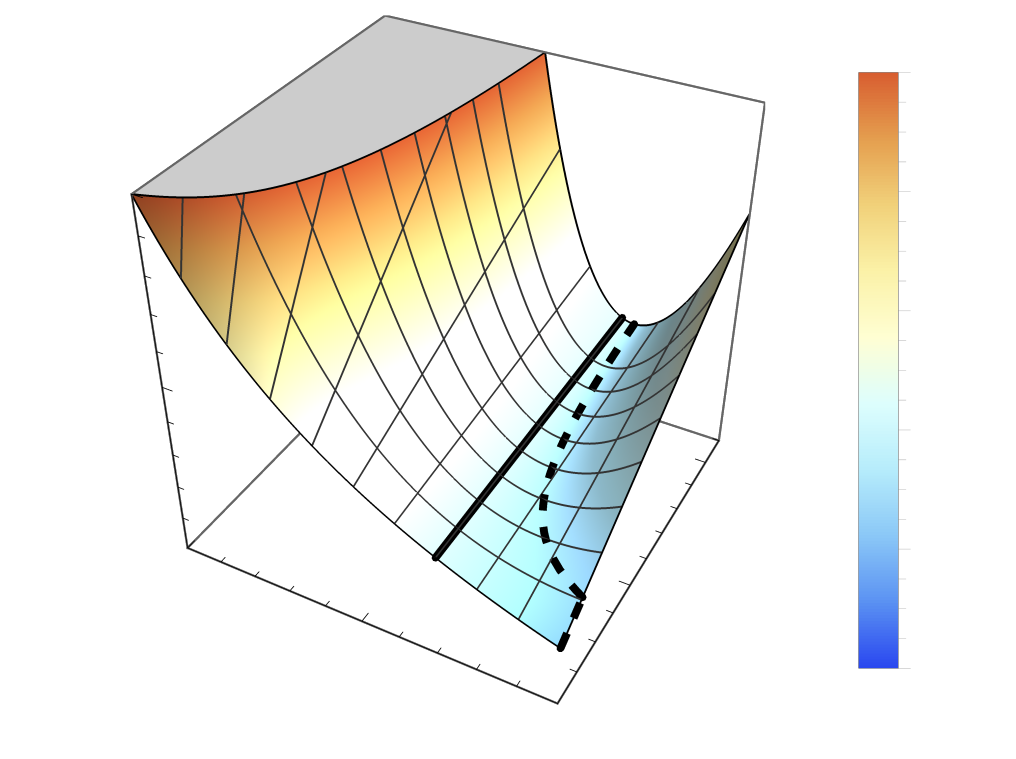
	\caption{The \gls{nrf} from Eq.~\ref{eqn:nrfSuperPoissonianNoNoise} scaling with relative channel efficiency (given $\eta_1=0.7$) is shown for values of $F$ from $1$ to $10$. The minimum \gls{nrf} is indicated by the dashed line, according to Eq.~\ref{eqn:superPoissonMinNRF}. The solid line indicates perfectly balanced detection, $\eta_2=\eta_1$, to show the convergence of the minimum \gls{nrf} for large $F$.
	}
	\label{fig:NRFtheoryDouble}
\end{figure}

We may write the lossy twin-beam variance in terms of the lossless photon number Fano factor $F$ using $\beta\mean{N}=F-1$:
\begin{equation}\label{eqn:superPoissonVar}
    \var{N_i}=\eta_i\mean{N}+\eta_i^2(F-1)\mean{N}.
\end{equation}
Using Eqs.~\ref{eqn:superPoissonVar} and \ref{eqn:beamMean} in Eq.~\ref{eqn:noiseReductionFactor} yields
\begin{equation}
    \sigma=\underbrace{1-\frac{2\eta_1\eta_2}{\eta_1+\eta_2}}_{\sigma_p}+    \underbrace{\frac{(\eta_1-\eta_2)^2(F-1)}{\eta_1+\eta_2}}_{\sigma_{sp}},
    \label{eqn:nrfSuperPoissonianNoNoise}
\end{equation}
where we use the covariance $\cov{N_1}{N_2}=\eta_1\eta_2(\mean{N}+\beta\mean{N}^2)$~\cite{Samantaray2017Thesis,Purcell1956,Mandel1958,Mandel1959}. The first two terms of Eq.~\ref{eqn:nrfSuperPoissonianNoNoise} ($\sigma_{p}$) correspond to correlated coherent-state intensity fluctuations. The final term ($\sigma_{sp}$) is the contribution to the \gls{nrf} associated with super-Poissonian intensity fluctuations. Importantly, $\sigma_{sp}$ has a dependence on the channel-efficiency mismatch and Fano factor; the minimum \gls{nrf} is therefore achieved when
\begin{equation}\label{eqn:superPoissonMinNRF}
    \eta_2=\begin{cases}
    1, & 1\leq F\leq F'\\
    \eta_1\left(\sqrt{4+\frac{2}{F-1}}-1\right), & F>F',
    \end{cases}
\end{equation}
where $F'=(\eta_1^2-2\eta_1-1)/(3\eta_1^2-2\eta_1-1)$. If $F$ is small, then increasing channel efficiency improves measured twin-beam correlations; if $F$ is large, then channel efficiency should approach balanced detection ($\eta_2\rightarrow\eta_1$) to improve measured correlations, as shown in Fig.~\ref{fig:NRFtheoryDouble}.

We now demonstrate the detrimental effects of uncorrelated noise photons and uncorrelated detector noise on the measured \gls{nrf}. Uncorrelated optical noise may come from resonant and non-resonant optical processes within the material used for twin-beam generation, such as fluorescence or broadband Raman scattering caused by the pump beam. Optical noise can also be associated with scattered or ambient light. Detector noise may be associated with photodiode dark current or CCD dark counts. The following derived model is general enough to account for all of these sources of noise. For simplicity, we include optical noise $N_\varrho$ on only one detection channel as $N_2\rightarrow N_2+N_\varrho$, and the procedure may be similarly adapted to accommodate independent noise on $N_1$ as well (see Appendix B). 

We define the optical noise to have the following realistic properties: (1) the efficiency for detecting the correlated-signal and uncorrelated-noise photons on channel $2$ is assumed to be the same: $\eta_\varrho=\eta_2$; (2) the mean noise intensity is some fraction $\varrho\geq0$ of the mean signal intensity: $\mean{N_\varrho}=\eta_2\varrho\mean{N}$; (3) the Fano factor of the optical noise $F_\varrho$ before channel loss can be written as $F_\varrho-1=\varrho(F-1)\geq0$, according to Eq.~\ref{eqn:superPoissonBeamVar}; and (4) the optical noise photons are generated via a process distinct from the signal photons: $\cov{N_1}{N_\varrho}=\cov{N_2}{N_\varrho}=0$. Similar to Eq.~\ref{eqn:superPoissonVar}, we have
\begin{equation}
\var{N_\varrho}=\eta_2\varrho\mean{N}+\eta_2^2\varrho(F_\varrho-1)\mean{N}.\label{eqn:ramanNRFequalities}
\end{equation}

When considering detector noise, we assume it to be the same for both channels, $N_{\{1,2\}}\rightarrow N_{\{1,2\}}+N_d$ (see Appendix B for unbalanced detector noise). Let the mean dark counts of each detector $\mean{N_d}$ be some fraction $d$ of the optical signal $\mean{N}$, independent of $\eta_1$ and $\eta_2$, with corresponding Fano factor $F_d$. Although detector noise cannot be explicitly derived from the optical signal, characterizing experimental parameters allows one to draw this equivalence. The variance of the detector noise may be written as 
\begin{equation}\label{eqn:detectorNRFequalities}
\var{N_d}=d\mean{N}+d(F_d-1)\mean{N}.
\end{equation}
We note the covariance for each detector with each other and the optical signals is zero, because they are uncorrelated. 

Finally, we can write the \gls{nrf} for individually super-Poissonian twin beams, accounting for uncorrelated, super-Poissonian optical and detector noise, combining the results of Eqs.~\ref{eqn:ramanNRFequalities} and \ref{eqn:detectorNRFequalities} in Eq.~\ref{eqn:noiseReductionFactor}: $\sigma=\sigma_p+\sigma_{sp}+\sigma_\varrho+\sigma_d$, with
\begin{subequations}
\begin{gather}
    \sigma_p=1-\frac{2\eta_1\eta_2}{\eta_1+(1+\varrho)\eta_2+2d}\label{eqn:NRFfullPoiss}\\[6pt]
    \sigma_{sp}=\frac{(\eta_1-\eta_2)^2(F-1)}{\eta_1+(1+\varrho)\eta_2+2d}\\[6pt]
    \sigma_\varrho=\frac{\eta_2^2\varrho(F_\varrho-1)}{\eta_1+(1+\varrho)\eta_2+2d}\\[6pt]
    \sigma_d=\frac{2d(F_d-1)}{\eta_1+(1+\varrho)\eta_2+2d}\label{eqn:NRFfulldetector},
\end{gather}
\end{subequations}
where $\sigma_\varrho$ and $\sigma_d$ are the optical and detector noise contributions to the \gls{nrf}, respectively.

Similar to before, we can now calculate the relative channel efficiency required to obtain the minimum \gls{nrf}. Assuming detector noise is much less than optical noise, $d\ll\varrho$, as may be the case for bright optical signals or low-noise detection, this minimum is achieved when
\begin{equation}\label{eqn:ramanNRFmin}
    \eta_2=\begin{cases}
    1, & 1\leq F\leq F''\\
   \frac{\eta_1}{1+\varrho}\left(\sqrt{2+\frac{2(2\varrho F+F-\varrho)}{(F-1)(1+\varrho^2)}}-1\right), & F>F'',
    \end{cases}
\end{equation}
where $F''=1-2\eta_1^2/[\eta_1^2(\varrho+3)-(2\eta_1+\varrho+1)(1+\varrho^2)]$. As optical noise increases in Eq.~\ref{eqn:ramanNRFmin}, $\eta_2\rightarrow0$, and $\sigma\rightarrow F$. This result highlights the importance of reducing optical noise on the twin beams to minimize the \gls{nrf}.

\begin{figure}
    \centering
	\def\svgwidth{0.8\columnwidth}
	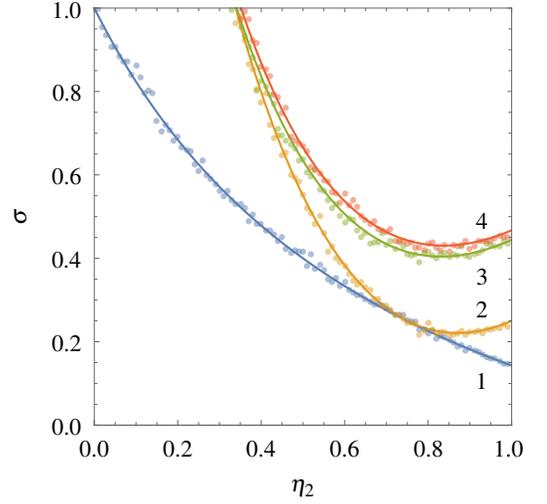
	\caption{The \gls{nrf} scaling with relative channel efficiency ($\eta_1=0.75$) is shown for realistic experimental parameters. 1: Poissonian twin beams ($F=1$); 2: super-Poissonian twin beams ($F=4$); 3: super-Poissonian twin beams with uncorrelated super-Poissonian optical noise ($\varrho=0.45$ and $F_\varrho=1.2$); 4: super-Poissonian twin beams with uncorrelated super-Poissonian optical and detector noise ($d=0.01$ and $F_d=3$). Correspondingly-colored points represent statistical simulation results, showing close agreement with theory (details in Appendix C). Note in plot 4, $\sigma$ equals Fano and detector noise for $\eta_2=0$ (see Eqs.~\ref{eqn:NRFfullPoiss}--\ref{eqn:NRFfulldetector}), and is therefore greater than one.
	}
	\label{fig:NRFcomparison}
\end{figure}

For specific values characterizing the optical signal and optical and detector noise, we can compare, as in Fig.~\ref{fig:NRFcomparison}, the behavior of the \gls{nrf} as relative channel efficiency is varied, from the ideal Poissonian twin-beam case to the full model including uncorrelated optical and detector noise (see Eqs.~\ref{eqn:NRFfullPoiss}--\ref{eqn:NRFfulldetector}). In the ideal case, one always measures sub-Poissonian correlations when including a quantum-correlated twin beam. Using realistic values for the experimental parameters of Eqs.~\ref{eqn:NRFfullPoiss}--\ref{eqn:NRFfulldetector}, as shown in Fig.~\ref{fig:NRFcomparison}, however, the minimum-attainable \gls{nrf} increases, and sub-Poissonian twin-beam correlations may not be measurable despite the underpinning non-classical correlations between the beams.

\section{\label{sec:nrfWithExperiment}Noise-reduction factor results applied to an example experimental scenario}

We now derive an example scenario for a proposed twin beam experiment based on stimulated \gls{fwm} to illustrate the effects of noise at various optical powers and how experimental parameters may be optimized. We make the following three assumptions: (1) twin-beam power increases exponentially with pump power $p$ as $\mean{N}\propto pe^{\lambda p}$, for some constant $\lambda>0$; (2) optical noise power increases linearly with pump power as $\mean{N_\varrho}\propto p$; and (3) mean detector noise increases linearly with \textit{e.g.} integration time or size, or temperature $w$, and has some constant readout noise, but is independent of $p$: $\mean{N_d}\propto w+\lambda$, for some other constant $\lambda>0$.

\begin{figure}
    \centering
	\def\svgwidth{0.8957\columnwidth}
	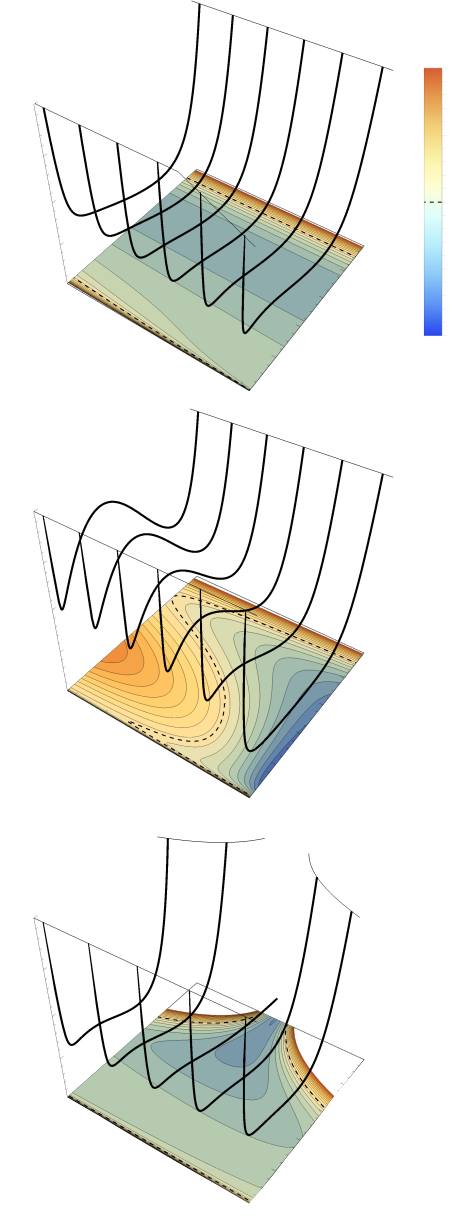
	\caption{The \gls{nrf} plotted against pump power for varying experimental parameters, with contour plots projected beneath: (a) increasing detector noise parameter $\lambda_8$, (b) increasing optical noise parameter $\lambda_6$, and (c) varying relative channel efficiency $\eta_2-\eta_1$ (see Eqs.~\ref{eqn:fanoPower}--\ref{eqn:dPower}). The default parameters for this simulation are $\Lambda=\{0.00005,0.01,0.01,0.5,0.1,1,0.005,1.001,0,0.005\}$, with $\eta_1=0.75$ and $\eta_2=0.7$. Dashed lines represent the $\sigma=1$ contour, and the scale bar is fixed for all plots.}
	\label{fig:nrfVaryParams}
\end{figure}

Assumption (1) is true if, for example, twin beams are generated via \gls{fwm} in the exponential-gain regime~\cite{Agrawal2012}. Assumption (2) may represent optical noise sources such as Raman scattering at wavelengths far from the pump, where the phonon density of states is low~\cite{Stolen1979,Aggarwal1970}. Assumption (3) represents thermal detector noise (linear contribution in $w$) and constant readout noise, typical for photodiode~\cite{Lvovsky2017} and CCD~\cite{Irie2008} detectors. Thus, we may write the terms of Eqs.~\ref{eqn:NRFfullPoiss}--\ref{eqn:NRFfulldetector} as
\begin{tabular*}{\textwidth}{ll}
\begin{minipeqn}
F(p)=1+\lambda_1pe^{\lambda_2p}\tag{12a}\label{eqn:fanoPower}
\end{minipeqn}&
\begin{minipeqn}
F_\varrho(p)=1+\lambda_3p\tag{12b}
\end{minipeqn}\\
\begin{minipeqn}
F_d(w)=1+\lambda_4+\lambda_5w\tag{12c}\label{eqn:fanoDetector}
\end{minipeqn}&
\begin{minipeqn}
\varrho(p)=\lambda_6e^{-\lambda_7p}\tag{12d}\label{eqn:rPower}
\end{minipeqn}
\end{tabular*}
\begin{equation}
d(p,w)=(\lambda_8w+\lambda_9)p^{-1}e^{-\lambda_{10}p},\tag{12e}\label{eqn:dPower}
\end{equation}
where $\Lambda=\{\lambda_1,\ldots,\lambda_{10}>0\}$ are fit parameters of the proposed experiment's model. Eqs.~\ref{eqn:fanoPower}--\ref{eqn:fanoDetector} are found by direct substitution of assumptions (1)--(3) into Eq.~\ref{eqn:superPoissonBeamVar}, with $\beta>0$; Eqs.~\ref{eqn:rPower} and \ref{eqn:dPower} by substitution into the definitions of $\varrho$ and $d$. Although detector noise is fundamentally independent of pump power, their relative intensities can still be compared using this substitution, as in Eq.~\ref{eqn:dPower}. 
The \gls{nrf} is plotted in Fig.~\ref{fig:nrfVaryParams} as a function of pump power, using Eqs.~\ref{eqn:NRFfullPoiss}--\ref{eqn:NRFfulldetector} and \ref{eqn:fanoPower}--\ref{eqn:dPower}, to show how various noise sources impact measured correlations.

In Fig.~\ref{fig:nrfVaryParams}~(a), detector noise is dominant at lower optical power, as the signal-to-noise ratio is low in this regime; indeed, as $p\rightarrow0$, then $\sigma\rightarrow F_d$, as expected. Increasing uncorrelated optical noise in Fig.~\ref{fig:nrfVaryParams}~(b) has a more significant effect at low and intermediate pump powers, highlighting the importance of filtering optical noise. 

As the signal-to-noise ratio increases in Fig.~\ref{fig:nrfVaryParams}~(c) at higher pump powers due to the different scalings of $\mean{N}$, $\mean{N_\varrho}$, and $\mean{N_d}$, balancing channel efficiency becomes the critical task. This is because the Fano factor of our considered example scales exponentially in pump power, so $\sigma_{sp}$ diverges quickly when $\eta_1\neq\eta_2$, as seen in Fig.~\ref{fig:NRFtheoryDouble}. Indeed, $\eta_2=0.8$ outperforms $\eta_2=1$ in Fig.~\ref{fig:nrfVaryParams}~(c), despite being lower channel efficiency. High-power intensity-correlation experiments should therefore implement classical noise suppression to reduce $F$~\cite{Allen2019}, or appropriately balance channel efficiency, to measure sub-Poissonian correlations.

\section{\label{sec:conclusion}Conclusions and outlook}

We have derived a novel model of twin-beam intensity correlations which accounts for experimental limitations such as uncorrelated optical and detector noise and unbalanced detection efficiency. From this model, we find that for beams with excess noise below a well-defined threshold, measured correlations can be improved by maximizing the detection efficiency of both beams. However, for beams with intensity or experimental noise beyond this threshold, one should appropriately balance detection efficiency, even at the cost of reducing channel efficiency.

We have also demonstrated the utility of this model for an example \gls{fwm} experimental scenario. While we have only considered this specific example, the model and the techniques used to derive it apply to many similar experiments, from wavelength-degenerate downconversion experiments dominated by detector noise, to non-wavelength-degenerate \gls{fwm} experiments dominated by optical noise~\cite{Moreau2017,Whittaker2017,Losero2018,Samantaray2017,Jakeman1986,Chesterking2017,Finger2015,Iskhakov2016}. 

Higher-power measurements find more application than intensities associated with squeezed vacuum and photon counting experiments. We therefore believe this model will find use as optical quantum metrology targets practical applications at microwatts to milliwatts of optical power and exotic wavelengths beyond the near-infrared, using high-gain twin-beam experiments to enhance measurement precision.

\section*{Acknowledgments and data availability}
We thank E.~Allen, M.~Chekhova, and J.~Rarity for helpful discussions. This work was supported by QuantIC -- The UK Quantum Technology Hub in Quantum Imaging, EPSRC Grant No. EP/T00097X/1.  J.~D.~M. was supported by the Quantum Engineering Centre for Doctoral Training, EPSRC Grant No. EP/L015730/1. J.~C.~F.~M.  acknowledges fellowship support from EPSRC Grant No. EP/M024385/1 and ERC starting Grant No. ERC-2018-STG803665. 

The data that support the findings of this study are available from the corresponding author upon reasonable request.

\section*{Appendix A: Comparison of twin-beam absorption estimators}

Estimators are mathematical formulas applied to finite data sets for determining physical parameters of a system. One such parameter used to characterize \textit{e.g.} biological samples is spectral absorption $\alpha(\lambda)$. Typically, for a given wavelength $\lambda$, measuring sample absorption involves comparing the intensity of a light source with and without a sample in its path: 
\begin{equation}\label{eqn:classicalAbsEstForm}
   \alpha_{c}=1-\frac{n'_1}{\mean{n_1}}, 
\end{equation}
where $0\leq\alpha_c\leq1$ is the direct classical absorption estimator, $n_1$ is the probe beam intensity for each measurement trial, and the prime denotes beam intensity after a lossy interaction with the sample. For the remainder of these discussions, $\alpha$ without a subscript refers to the population estimate (unbiased estimate based on an infinitely-large data set), and with refers to sample estimates (realistic finite-sized data sets).

The precision of this absorption measurement is limited by the Poissonian quantum nature of light, the \gls{snl}, as 
\begin{equation}\label{eqn:classicalAbsEst}
    \var{\alpha_c}=\frac{(1-\alpha)}{\mean{n_1}}.
\end{equation}

A first example of a twin-beam absorption estimator for quantum parameter estimation was presented in Ref.~\onlinecite{Jakeman1986} and further explored by Ref.~\onlinecite{Losero2018}:
\begin{equation}
    \alpha_l=1-\gamma\frac{n'_1}{n'_2},
\end{equation}
where $n_2$ is the reference beam intensity, and $\gamma=\mean{n_2}/\mean{n_1}$ accounts for unbalanced channel efficiency. Primes in this case denote the measurement stage in general, and the sample is only placed in the path of the probe beam $n_1$.

In the case of balanced channel efficiency ($\gamma=1$) and no optical or detector noise, one may write 
\begin{equation}\label{eqn:loseroAbsEst}
    \var{\alpha_l}=\var{\alpha_u}+2\frac{(1-\alpha)^2}{\mean{n_1}}\sigma^*,
\end{equation}
where $\var{\alpha_u}=\alpha\var{\alpha_c}$ is the ultimate quantum limit of an absorption measurement, associated with binomial measurement statistics, attainable with \textit{e.g.} Fock states or when $\sigma=0$~\cite{Whittaker2017,Losero2018}, and $\sigma^*=1-\eta$ is the noiseless, balanced-detection \gls{nrf}. To compare this twin-beam estimator to the classical direct case, we use their relative estimator efficiency
\begin{subequations}
\begin{align}
    \Gamma_i&=\frac{\mse{\alpha_i}}{\mse{\alpha_{c}}}\\
    &=\frac{\var{\alpha_i}+(\mean{\alpha_i}-\alpha)^2}{\var{\alpha_c}},
\end{align}
\end{subequations}
for some estimator $i$, where $\mse{\alpha_i}$ is the mean squared error, which equals $\var{\alpha_i}$ in the case of unbiased parameter estimation (as implicitly assumed in Refs.~\onlinecite{Moreau2017,Losero2018}). When $0\leq\Gamma_i<1$, the estimator efficiency is sub-\gls{snl}. This regime is exclusive to quantum-correlated twin beams, similar to $0\leq\sigma<1$. 

Comparing Eqs.~\ref{eqn:classicalAbsEst} and \ref{eqn:loseroAbsEst} yields
\begin{equation}
    \Gamma_l=\alpha+2(1-\alpha)\sigma^*.   
\end{equation}
One finds $\Gamma_l>1$ for all $\sigma^*>0.5$. Thus, even though beams may display sub-Poissonian intensity correlations, one cannot always perform sub-\gls{snl} absorption measurements with this estimator. One can gain insight into this counter-intuitive result by considering how $\alpha_c$ is an even less suitable estimator for the twin-beam case, as $\Gamma_c=1$ for all values of $\sigma$.

Ref.~\onlinecite{Moreau2017} presents another twin-beam absorption estimator:
\begin{equation}\label{eqn:MoreauAbsEst}
    \alpha_m=1-\frac{n'_1-k\delta n'_2+\delta E}{\mean{n_1}},
\end{equation}
where $\delta n'_2=n'_2-\mean{n'_2}$, $k$ is a weight factor used to maximize the estimator's precision, and $\delta E=\mean{k\delta n'_2}$ is a correction factor used to ensure that the estimator is unbiased (\textit{i.e.} $\mean{\alpha_m}=\alpha$). Contrary to Refs.~\onlinecite{Losero2018,Moreau2017}, $\alpha_m$ is indeed biased in the presence of classical intensity fluctuations, as we demonstrate at the end of this section. We also correct the estimator to be unbiased.

\begin{figure*}
	\centering
	\def\svgwidth{0.8\textwidth}
	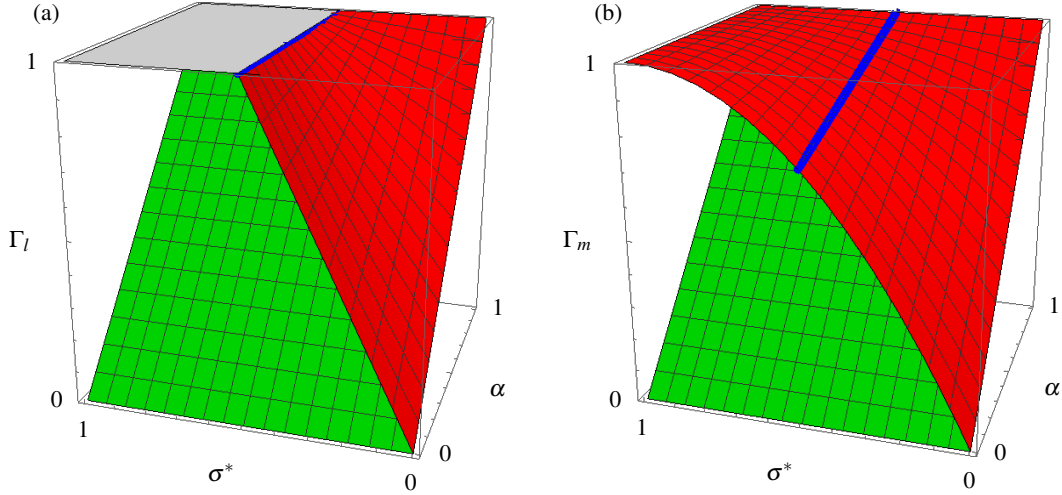
	\caption{Comparing relative \gls{snl} performance metrics (a) $\Gamma_l$ and (b) $\Gamma_m$ in the case of balanced channel efficiency and no optical or detector noise. The green plane $\Gamma_u=\alpha$ is the ultimate quantum limit, and the blue line is the $\sigma^*=0.5$ contour.}
	\label{fig:gammaSigma}
\end{figure*}

One may perform a similar analysis as the previous estimator, now with~\cite{Losero2018}
\begin{equation}
    \var{\alpha_m}=\var{\alpha_u}+2\frac{(1-\alpha)^2}{\mean{n'_1}}\sigma^*(1-\frac{\sigma^*}{2}),
\end{equation}
in the noiseless, balanced-detection case with optimized $k$~\cite{Moreau2017}:
\begin{equation}
    k_m^\textrm{opt}=\frac{\cov{n'_1}{n'_2}}{\var{n'_2}}.
\end{equation}
Comparing this to the classical direct measurement with $\gamma=1$,
\begin{equation}
    \Gamma_m=\alpha+2(1-\alpha)\sigma^*(1-\frac{\sigma^*}{2}).
\end{equation}
We now find sub-\gls{snl} $\Gamma_m$ for all $\sigma^*<1$, and $\Gamma_m<\Gamma_l$ for all $\sigma^*>0$ and $\alpha<1$. The performance of $\alpha_m$ and $\alpha_l$ is compared graphically in Fig.~\ref{fig:gammaSigma}. We see in this figure that $\alpha_m$ is a superior estimator to $\alpha_l$ when appropriately calibrated. In the case discussed here, one achieves sub-\gls{snl} measurement statistics for any values of $\eta_{\{1,2\}}>0$ using $\alpha_m$, relaxing the requirement that $\eta_{\{1,2\}}>0.5$ when using $\alpha_l$, stated in Ref.~\onlinecite{Jakeman1986}.

Although we do not derive it here, we expect from our discussions of the \gls{nrf} in the main text that super-Poissonian intensity noise with unbalanced channel efficiency and other uncorrelated noise sources further reduce the efficacy of $\alpha_l$ and $\alpha_m$ for achieving sub-\gls{snl} measurement statistics.

We will also show that twin-beam estimators are not only more precise than the direct classical absorption estimator, but also more accurate in general.

For stationary processes (processes whose mean and variance do not change with time), $\alpha_c$ is indeed unbiased, as $\mean{n'_1}=(1-\alpha)\mean{n_1}$, and $\mean{\alpha_c}=\alpha$. For non-stationary processes, however, the probe and reference beam powers are changed by an amount $\varepsilon\geq-1$:
\begin{gather}
    \mean{n'_1}=(1-\alpha)(1+\varepsilon)\mean{n_1}\label{eqn:biasedn1}\\[6pt]
    \mean{n'_2}=(1+\varepsilon)\mean{n_2}.\label{eqn:biasedn2}
\end{gather}
This may occur experimentally if the probe beam power is changed between the calibration and measurement phases. Because $\alpha_c$ does not have access to the reference beam, substitution of Eq.~\ref{eqn:biasedn1} into Eq.~\ref{eqn:classicalAbsEstForm} yields 
\begin{equation}\label{eqn:classicalBias}
    \mean{\alpha_c}=1-(1-\alpha)(1+\varepsilon),
\end{equation}
which is biased without knowledge of $\varepsilon$. Simply, the direct classical absorption estimator cannot distinguish probe beam intensity fluctuations from sample absorption.

Considering now the twin-beam estimator $\alpha_l$, we may substitute Eqs.~\ref{eqn:biasedn1}~and~\ref{eqn:biasedn2}, yielding
\begin{subequations}
\begin{align}
    \mean{\alpha_l}&=1-\gamma\mean{\frac{n'_1}{n'_2}}\\
    &\approx1-\frac{\mean{n_2}}{\mean{n_1}}\frac{\mean{n'_1}}{\mean{n'_2}}\\
    &=\alpha,
\end{align}
\end{subequations}
where the approximation in line two is valid for large $n$~\cite{Jakeman1986,Kempen2000}. This estimator is therefore unbiased in the large-photon-flux limit, which is the regime where intensity-correlated measurements are most practical.

Finally, we consider the absorption estimator $\alpha_m$, which we previously showed to obtain the greatest measurement precision of the three discussed estimators. The form of this estimator, as originally presented in Ref.~\onlinecite{Moreau2017} and discussed further in Ref.~\onlinecite{Losero2018}, is biased, obtaining the same functional form for $\mean{\alpha_m}$ as Eq.~\ref{eqn:classicalBias}: 
\begin{subequations}
\begin{align}
    \mean{\alpha_m}&\approx1-\frac{\mean{n'_1}-\mean{k\delta n'_2}+\delta E}{\mean{n_1}}\\
    &=1-\frac{\mean{n'_1}}{\mean{n_1}}\\
    &=1-(1-\alpha)(1+\varepsilon).
\end{align}
\end{subequations}
This is because $\alpha_m$ is derived from $\alpha_c$, which implicitly requires a stationary twin-beam intensity to be unbiased. We present here an new, unbiased form of $\alpha_m$, denoted $\alpha_{lm}$, using $\alpha_l$ as the starting point:
\begin{equation}
    \alpha_{lm}=1-\gamma\frac{n'_1-k\delta n'_2+\delta E}{n'_2}.
\end{equation}
This estimator is unbiased for optimized $k$, as $\mean{\alpha_{lm}}=\mean{\alpha_l}=\alpha$.

The $k$ which maximizes the precision of $\alpha_{lm}$ is found by minimizing $\var{(n'_1-k\delta n'_2)/n'_2}$. This variance may be approximated according to Ref.~\onlinecite{Kempen2000}, yielding
\begin{equation}
    k_{lm}^\textrm{opt}\approx k_m^\textrm{opt}-\frac{\mean{n'_1}}{\mean{n'_2}}.
\end{equation}

\section*{Appendix B: Noise-reduction factor with uncorrelated noise on both detection channels}

We derived the \gls{nrf} Eqs.~\ref{eqn:NRFfullPoiss}--\ref{eqn:NRFfulldetector} for the case of optical noise on only one detection channel and balanced detector noise on both channels, for simplicity. These equations may be generalized to include uncorrelated optical and detection noise on each channel following the same procedure outlined in the main text, with the following result:
\begingroup
\allowdisplaybreaks
\begin{subequations}
\begin{align}
    \sigma_p&=1-\frac{2\eta_1\eta_2}{(1+\varrho_1)\eta_1+(1+\varrho_2)\eta_2+d_1+d_2}\\[6pt]
    \sigma_{sp}&=\frac{(\eta_1-\eta_2)^2(F-1)}{(1+\varrho_1)\eta_1+(1+\varrho_2)\eta_2+d_1+d_2}\\[6pt]
    \sigma_\varrho&=\frac{\eta_1^2\varrho_1(F_{\varrho_1}-1)+\eta_2^2\varrho_2(F_{\varrho_2}-1)}{(1+\varrho_1)\eta_1+(1+\varrho_2)\eta_2+d_1+d_2}\\[6pt]
    \sigma_d&=\frac{d_1(F_{d_1}-1)+d_2(F_{d_2}-1)}{(1+\varrho_1)\eta_1+(1+\varrho_2)\eta_2+d_1+d_2},
\end{align}
\end{subequations}
\endgroup
where $N_{\{1,2\}}\rightarrow N_{\{1,2\}}+N_{\varrho_{\{1,2\}}}+N_{d_{\{1,2\}}}$. Setting $\varrho_1=0$, $d_1=d_2$, and $F_{d_1}=F_{d_2}$ yields the derived Eqs.~\ref{eqn:NRFfullPoiss}--\ref{eqn:NRFfulldetector}.

\newpage

\section*{Appendix C: Details of noise-reduction factor simulation for experimental model}

The simulations shown in Fig.~\ref{fig:NRFcomparison} were performed according to the following procedure.

We first define the mean and variance the distributions $N$, $N_\varrho$, and $N_d$ from which the signal counts and optical and detector noise counts are sampled. These distributions are Gaussian for large mean values, where the degree to which they are super-Poissonian can be set by the relative values of their means and variances. We also define the number of trials $t$ for the data to be averaged over, as well as channel detection efficiency $\eta_1$. 

For each count source (twin beams, optical noise, and detector noise), an integer list of length $t$ is generated, with each element sampled from its corresponding distribution. This represents the number of pre-loss photons or detector dark counts, for each measurement trial.

A loop is performed over $\eta_2$ from $0$ to $1$. Within this loop, a loop over $t$ is performed, where for each trial and each count source, a list of pseudo-random numbers between $0$ to $1$, inclusive, is generated whose length is given according to the the specified element from the previous step. To determine if the photon is detected as a count, these pseudo-random numbers are compared to the correspondingly defined channel efficiency, and replaced with a one if the pseudo-random number is less than $\eta_{\{1,2\}}$, zero otherwise (detector noise counts, independent of detector efficiency, do not undergo this comparison). The list is then summed and stored as the number of detected counts for that trial. In this way, we can simulate the random loss associated with the photon-count sources.

Finally, the signal and noise counts are summed for each channel, the \gls{nrf} is calculated for the specified $\eta_2$, and $\eta_2$ is incremented.

\bibliography{biblio}

\end{document}

%% file: figs/NRFcontourPlots.pdf_tex
%% Creator: Inkscape inkscape 0.92.4, www.inkscape.org
%% PDF/EPS/PS + LaTeX output extension by Johan Engelen, 2010
%% Accompanies image file 'NRFcontourPlots.pdf' (pdf, eps, ps)
%%
%% To include the image in your LaTeX document, write
%%   \input{<filename>.pdf_tex}
%%  instead of
%%   \includegraphics{<filename>.pdf}
%% To scale the image, write
%%   \def\svgwidth{<desired width>}
%%   \input{<filename>.pdf_tex}
%%  instead of
%%   \includegraphics[width=<desired width>]{<filename>.pdf}
%%
%% Images with a different path to the parent latex file can
%% be accessed with the `import' package (which may need to be
%% installed) using
%%   \usepackage{import}
%% in the preamble, and then including the image with
%%   \import{<path to file>}{<filename>.pdf_tex}
%% Alternatively, one can specify
%%   \graphicspath{{<path to file>/}}
%% 
%% For more information, please see info/svg-inkscape on CTAN:
%%   http://tug.ctan.org/tex-archive/info/svg-inkscape
%%
\begingroup%
  \makeatletter%
  \providecommand\color[2][]{%
    \errmessage{(Inkscape) Color is used for the text in Inkscape, but the package 'color.sty' is not loaded}%
    \renewcommand\color[2][]{}%
  }%
  \providecommand\transparent[1]{%
    \errmessage{(Inkscape) Transparency is used (non-zero) for the text in Inkscape, but the package 'transparent.sty' is not loaded}%
    \renewcommand\transparent[1]{}%
  }%
  \providecommand\rotatebox[2]{#2}%
  \newcommand*\fsize{\dimexpr\f@size pt\relax}%
  \newcommand*\lineheight[1]{\fontsize{\fsize}{#1\fsize}\selectfont}%
  \ifx\svgwidth\undefined%
    \setlength{\unitlength}{244.00007593bp}%
    \ifx\svgscale\undefined%
      \relax%
    \else%
      \setlength{\unitlength}{\unitlength * \real{\svgscale}}%
    \fi%
  \else%
    \setlength{\unitlength}{\svgwidth}%
  \fi%
  \global\let\svgwidth\undefined%
  \global\let\svgscale\undefined%
  \makeatother%
  \begin{picture}(1,0.7407252)%
    \lineheight{1}%
    \setlength\tabcolsep{0pt}%
    \put(0,0){\includegraphics[width=\unitlength,page=1]{NRFcontourPlots.png}}%
    \put(0.90566443,0.07525179){\color[rgb]{0,0,0}\makebox(0,0)[lt]{\lineheight{1.25}\smash{\begin{tabular}[t]{l}0.0\end{tabular}}}}%
    \put(0.90566443,0.19173083){\color[rgb]{0,0,0}\makebox(0,0)[lt]{\lineheight{1.25}\smash{\begin{tabular}[t]{l}0.2\end{tabular}}}}%
    \put(0.90566443,0.54367743){\color[rgb]{0,0,0}\makebox(0,0)[lt]{\lineheight{1.25}\smash{\begin{tabular}[t]{l}0.8\end{tabular}}}}%
    \put(0.90566443,0.30853407){\color[rgb]{0,0,0}\makebox(0,0)[lt]{\lineheight{1.25}\smash{\begin{tabular}[t]{l}0.4\end{tabular}}}}%
    \put(0.90201033,0.66048067){\color[rgb]{0,0,0}\makebox(0,0)[lt]{\lineheight{1.25}\smash{\begin{tabular}[t]{l}1.0\end{tabular}}}}%
    \put(0.90566443,0.42610575){\color[rgb]{0,0,0}\makebox(0,0)[lt]{\lineheight{1.25}\smash{\begin{tabular}[t]{l}0.6\end{tabular}}}}%
    \put(0.30628333,0.05531481){\color[rgb]{0,0,0}\makebox(0,0)[t]{\lineheight{1.25}\smash{\begin{tabular}[t]{c}$\eta_2$\end{tabular}}}}%
    \put(0.1717942,0.16147895){\color[rgb]{0,0,0}\makebox(0,0)[t]{\lineheight{1.25}\smash{\begin{tabular}[t]{c}0.0\end{tabular}}}}%
    \put(0.34389498,0.08858997){\color[rgb]{0,0,0}\makebox(0,0)[t]{\lineheight{1.25}\smash{\begin{tabular}[t]{c}0.5\end{tabular}}}}%
    \put(0.53860819,0.00676602){\color[rgb]{0,0,0}\makebox(0,0)[t]{\lineheight{1.25}\smash{\begin{tabular}[t]{c}1.0\end{tabular}}}}%
    \put(0.68806973,0.1432851){\color[rgb]{0,0,0}\makebox(0,0)[lt]{\lineheight{1.25}\smash{\begin{tabular}[t]{l}$F$\end{tabular}}}}%
    \put(0.5895851,0.06020285){\color[rgb]{0,0,0}\makebox(0,0)[lt]{\lineheight{1.25}\smash{\begin{tabular}[t]{l}2\end{tabular}}}}%
    \put(0.63050841,0.12384781){\color[rgb]{0,0,0}\makebox(0,0)[lt]{\lineheight{1.25}\smash{\begin{tabular}[t]{l}4\end{tabular}}}}%
    \put(0.66425202,0.18266377){\color[rgb]{0,0,0}\makebox(0,0)[lt]{\lineheight{1.25}\smash{\begin{tabular}[t]{l}6\end{tabular}}}}%
    \put(0.69786375,0.23636424){\color[rgb]{0,0,0}\makebox(0,0)[lt]{\lineheight{1.25}\smash{\begin{tabular}[t]{l}8\end{tabular}}}}%
    \put(0.71511974,0.28378526){\color[rgb]{0,0,0}\makebox(0,0)[lt]{\lineheight{1.25}\smash{\begin{tabular}[t]{l}10\end{tabular}}}}%
    \put(0.07190431,0.37623051){\color[rgb]{0,0,0}\rotatebox{90}{\makebox(0,0)[t]{\lineheight{1.25}\smash{\begin{tabular}[t]{c}$\sigma$\end{tabular}}}}}%
    \put(0.15091279,0.35521341){\color[rgb]{0,0,0}\makebox(0,0)[rt]{\lineheight{1.25}\smash{\begin{tabular}[t]{r}0.5\end{tabular}}}}%
    \put(0.11849696,0.54670493){\color[rgb]{0,0,0}\makebox(0,0)[rt]{\lineheight{1.25}\smash{\begin{tabular}[t]{r}1.0\end{tabular}}}}%
  \end{picture}%
\endgroup%

%% file: figs/NRFcomparisonWithSims.pdf_tex
%% Creator: Inkscape inkscape 0.92.4, www.inkscape.org
%% PDF/EPS/PS + LaTeX output extension by Johan Engelen, 2010
%% Accompanies image file 'NRFcomparisonWithSims.pdf' (pdf, eps, ps)
%%
%% To include the image in your LaTeX document, write
%%   \input{<filename>.pdf_tex}
%%  instead of
%%   \includegraphics{<filename>.pdf}
%% To scale the image, write
%%   \def\svgwidth{<desired width>}
%%   \input{<filename>.pdf_tex}
%%  instead of
%%   \includegraphics[width=<desired width>]{<filename>.pdf}
%%
%% Images with a different path to the parent latex file can
%% be accessed with the `import' package (which may need to be
%% installed) using
%%   \usepackage{import}
%% in the preamble, and then including the image with
%%   \import{<path to file>}{<filename>.pdf_tex}
%% Alternatively, one can specify
%%   \graphicspath{{<path to file>/}}
%% 
%% For more information, please see info/svg-inkscape on CTAN:
%%   http://tug.ctan.org/tex-archive/info/svg-inkscape
%%
\begingroup%
  \makeatletter%
  \providecommand\color[2][]{%
    \errmessage{(Inkscape) Color is used for the text in Inkscape, but the package 'color.sty' is not loaded}%
    \renewcommand\color[2][]{}%
  }%
  \providecommand\transparent[1]{%
    \errmessage{(Inkscape) Transparency is used (non-zero) for the text in Inkscape, but the package 'transparent.sty' is not loaded}%
    \renewcommand\transparent[1]{}%
  }%
  \providecommand\rotatebox[2]{#2}%
  \newcommand*\fsize{\dimexpr\f@size pt\relax}%
  \newcommand*\lineheight[1]{\fontsize{\fsize}{#1\fsize}\selectfont}%
  \ifx\svgwidth\undefined%
    \setlength{\unitlength}{393.88424683bp}%
    \ifx\svgscale\undefined%
      \relax%
    \else%
      \setlength{\unitlength}{\unitlength * \real{\svgscale}}%
    \fi%
  \else%
    \setlength{\unitlength}{\svgwidth}%
  \fi%
  \global\let\svgwidth\undefined%
  \global\let\svgscale\undefined%
  \makeatother%
  \begin{picture}(1,0.96168105)%
    \lineheight{1}%
    \setlength\tabcolsep{0pt}%
    \put(0,0){\includegraphics[width=\unitlength,page=1]{NRFcomparisonWithSims.pdf}}%
    \put(0.17109122,0.07956616){\color[rgb]{0,0,0}\makebox(0,0)[t]{\lineheight{1.25}\smash{\begin{tabular}[t]{c}\small 0.0\end{tabular}}}}%
    \put(0.33170607,0.07956616){\color[rgb]{0,0,0}\makebox(0,0)[t]{\lineheight{1.25}\smash{\begin{tabular}[t]{c}\small 0.2\end{tabular}}}}%
    \put(0.49196639,0.07956616){\color[rgb]{0,0,0}\makebox(0,0)[t]{\lineheight{1.25}\smash{\begin{tabular}[t]{c}\small 0.4\end{tabular}}}}%
    \put(0.65253661,0.07956616){\color[rgb]{0,0,0}\makebox(0,0)[t]{\lineheight{1.25}\smash{\begin{tabular}[t]{c}\small 0.6\end{tabular}}}}%
    \put(0.81351967,0.07956616){\color[rgb]{0,0,0}\makebox(0,0)[t]{\lineheight{1.25}\smash{\begin{tabular}[t]{c}\small 0.8\end{tabular}}}}%
    \put(0.96708215,0.07956616){\color[rgb]{0,0,0}\makebox(0,0)[t]{\lineheight{1.25}\smash{\begin{tabular}[t]{c}\small 1.0\end{tabular}}}}%
    \put(0.14643817,0.12166978){\color[rgb]{0,0,0}\makebox(0,0)[rt]{\lineheight{1.25}\smash{\begin{tabular}[t]{r}\small 0.0\end{tabular}}}}%
    \put(0.14676544,0.28212103){\color[rgb]{0,0,0}\makebox(0,0)[rt]{\lineheight{1.25}\smash{\begin{tabular}[t]{r}\small 0.2\end{tabular}}}}%
    \put(0.14638361,0.44257232){\color[rgb]{0,0,0}\makebox(0,0)[rt]{\lineheight{1.25}\smash{\begin{tabular}[t]{r}\small 0.4\end{tabular}}}}%
    \put(0.1466018,0.60303356){\color[rgb]{0,0,0}\makebox(0,0)[rt]{\lineheight{1.25}\smash{\begin{tabular}[t]{r}\small 0.6\end{tabular}}}}%
    \put(0.14766543,0.76348477){\color[rgb]{0,0,0}\makebox(0,0)[rt]{\lineheight{1.25}\smash{\begin{tabular}[t]{r}\small 0.8\end{tabular}}}}%
    \put(0.14643817,0.92393598){\color[rgb]{0,0,0}\makebox(0,0)[rt]{\lineheight{1.25}\smash{\begin{tabular}[t]{r}\small 1.0\end{tabular}}}}%
    \put(0.57216971,0.01205445){\color[rgb]{0,0,0}\makebox(0,0)[t]{\lineheight{1.25}\smash{\begin{tabular}[t]{c}\small $\eta_2$\end{tabular}}}}%
    \put(0.0400905,0.54122544){\color[rgb]{0,0,0}\rotatebox{90}{\makebox(0,0)[t]{\lineheight{1.25}\smash{\begin{tabular}[t]{c}\small $\sigma$\end{tabular}}}}}%
    \put(0.91585682,0.21033901){\color[rgb]{0,0,0}\makebox(0,0)[t]{\lineheight{1.25}\smash{\begin{tabular}[t]{c}\small 1\end{tabular}}}}%
    \put(0.91627955,0.34681007){\color[rgb]{0,0,0}\makebox(0,0)[t]{\lineheight{1.25}\smash{\begin{tabular}[t]{c}\small 2\end{tabular}}}}%
    \put(0.91690681,0.41094499){\color[rgb]{0,0,0}\makebox(0,0)[t]{\lineheight{1.25}\smash{\begin{tabular}[t]{c}\small 3\end{tabular}}}}%
    \put(0.91625227,0.5171685){\color[rgb]{0,0,0}\makebox(0,0)[t]{\lineheight{1.25}\smash{\begin{tabular}[t]{c}\small 4\end{tabular}}}}%
  \end{picture}%
\endgroup%

%% file: figs/varyParams3D_2OSA__pcLight2.pdf_tex
%% Creator: Inkscape inkscape 0.92.4, www.inkscape.org
%% PDF/EPS/PS + LaTeX output extension by Johan Engelen, 2010
%% Accompanies image file 'varyParams3D(2OSA)_pcLight2.pdf' (pdf, eps, ps)
%%
%% To include the image in your LaTeX document, write
%%   \input{<filename>.pdf_tex}
%%  instead of
%%   \includegraphics{<filename>.pdf}
%% To scale the image, write
%%   \def\svgwidth{<desired width>}
%%   \input{<filename>.pdf_tex}
%%  instead of
%%   \includegraphics[width=<desired width>]{<filename>.pdf}
%%
%% Images with a different path to the parent latex file can
%% be accessed with the `import' package (which may need to be
%% installed) using
%%   \usepackage{import}
%% in the preamble, and then including the image with
%%   \import{<path to file>}{<filename>.pdf_tex}
%% Alternatively, one can specify
%%   \graphicspath{{<path to file>/}}
%% 
%% For more information, please see info/svg-inkscape on CTAN:
%%   http://tug.ctan.org/tex-archive/info/svg-inkscape
%%
\begingroup%
  \makeatletter%
  \providecommand\color[2][]{%
    \errmessage{(Inkscape) Color is used for the text in Inkscape, but the package 'color.sty' is not loaded}%
    \renewcommand\color[2][]{}%
  }%
  \providecommand\transparent[1]{%
    \errmessage{(Inkscape) Transparency is used (non-zero) for the text in Inkscape, but the package 'transparent.sty' is not loaded}%
    \renewcommand\transparent[1]{}%
  }%
  \providecommand\rotatebox[2]{#2}%
  \newcommand*\fsize{\dimexpr\f@size pt\relax}%
  \newcommand*\lineheight[1]{\fontsize{\fsize}{#1\fsize}\selectfont}%
  \ifx\svgwidth\undefined%
    \setlength{\unitlength}{351.93523497bp}%
    \ifx\svgscale\undefined%
      \relax%
    \else%
      \setlength{\unitlength}{\unitlength * \real{\svgscale}}%
    \fi%
  \else%
    \setlength{\unitlength}{\svgwidth}%
  \fi%
  \global\let\svgwidth\undefined%
  \global\let\svgscale\undefined%
  \makeatother%
  \begin{picture}(1,2.60522041)%
    \lineheight{1}%
    \setlength\tabcolsep{0pt}%
    \put(0,0){\includegraphics[width=\unitlength,page=1]{varyParams3D_2OSA__pcLight2.png}}%
    \put(0.70543852,1.86870441){\color[rgb]{0,0,0}\makebox(0,0)[lt]{\lineheight{1.25}\smash{\begin{tabular}[t]{l}$p$ (a.u.)\end{tabular}}}}%
    \put(0.53988496,1.73527906){\color[rgb]{0,0,0}\makebox(0,0)[lt]{\lineheight{1.25}\smash{\begin{tabular}[t]{l}0\end{tabular}}}}%
    \put(0.59718371,1.80743872){\color[rgb]{0,0,0}\makebox(0,0)[lt]{\lineheight{1.25}\smash{\begin{tabular}[t]{l}2\end{tabular}}}}%
    \put(0.64915396,1.87238321){\color[rgb]{0,0,0}\makebox(0,0)[lt]{\lineheight{1.25}\smash{\begin{tabular}[t]{l}4\end{tabular}}}}%
    \put(0.69745757,1.93478131){\color[rgb]{0,0,0}\makebox(0,0)[lt]{\lineheight{1.25}\smash{\begin{tabular}[t]{l}6\end{tabular}}}}%
    \put(0.74281401,1.99145904){\color[rgb]{0,0,0}\makebox(0,0)[lt]{\lineheight{1.25}\smash{\begin{tabular}[t]{l}8\end{tabular}}}}%
    \put(0.78280188,2.04605986){\color[rgb]{0,0,0}\makebox(0,0)[lt]{\lineheight{1.25}\smash{\begin{tabular}[t]{l}10\end{tabular}}}}%
    \put(0.02243456,2.16937908){\color[rgb]{0,0,0}\rotatebox{90}{\makebox(0,0)[t]{\lineheight{1.25}\smash{\begin{tabular}[t]{c}$\sigma$\end{tabular}}}}}%
    \put(0.1234414,1.98651792){\color[rgb]{0,0,0}\makebox(0,0)[rt]{\lineheight{1.25}\smash{\begin{tabular}[t]{r}0.0\end{tabular}}}}%
    \put(0.10806351,2.07126352){\color[rgb]{0,0,0}\makebox(0,0)[rt]{\lineheight{1.25}\smash{\begin{tabular}[t]{r}0.5\end{tabular}}}}%
    \put(0.09044857,2.16053348){\color[rgb]{0,0,0}\makebox(0,0)[rt]{\lineheight{1.25}\smash{\begin{tabular}[t]{r}1.0\end{tabular}}}}%
    \put(0.07287714,2.2589766){\color[rgb]{0,0,0}\makebox(0,0)[rt]{\lineheight{1.25}\smash{\begin{tabular}[t]{r}1.5\end{tabular}}}}%
    \put(0.0511959,2.3718276){\color[rgb]{0,0,0}\makebox(0,0)[rt]{\lineheight{1.25}\smash{\begin{tabular}[t]{r}2.0\end{tabular}}}}%
    \put(0.2871099,1.81897809){\color[rgb]{0,0,0}\makebox(0,0)[rt]{\lineheight{1.25}\smash{\begin{tabular}[t]{r}$\lambda_8$\end{tabular}}}}%
    \put(0.50047055,1.73781833){\color[rgb]{0,0,0}\makebox(0,0)[rt]{\lineheight{1.25}\smash{\begin{tabular}[t]{r}2\end{tabular}}}}%
    \put(0.42238257,1.78428653){\color[rgb]{0,0,0}\makebox(0,0)[rt]{\lineheight{1.25}\smash{\begin{tabular}[t]{r}4\end{tabular}}}}%
    \put(0.34781105,1.82712546){\color[rgb]{0,0,0}\makebox(0,0)[rt]{\lineheight{1.25}\smash{\begin{tabular}[t]{r}6\end{tabular}}}}%
    \put(0.27840801,1.86724533){\color[rgb]{0,0,0}\makebox(0,0)[rt]{\lineheight{1.25}\smash{\begin{tabular}[t]{r}8\end{tabular}}}}%
    \put(0.21061276,1.90829863){\color[rgb]{0,0,0}\makebox(0,0)[rt]{\lineheight{1.25}\smash{\begin{tabular}[t]{r}10\end{tabular}}}}%
    \put(0.14466988,1.9466237){\color[rgb]{0,0,0}\makebox(0,0)[rt]{\lineheight{1.25}\smash{\begin{tabular}[t]{r}12\end{tabular}}}}%
    \put(0.20841922,2.46114503){\color[rgb]{0,0,0}\makebox(0,0)[lt]{\lineheight{1.25}\smash{\begin{tabular}[t]{l}(a)\end{tabular}}}}%
    \put(0.70543852,1.00232894){\color[rgb]{0,0,0}\makebox(0,0)[lt]{\lineheight{1.25}\smash{\begin{tabular}[t]{l}$p$ (a.u.)\end{tabular}}}}%
    \put(0.53988496,0.86890359){\color[rgb]{0,0,0}\makebox(0,0)[lt]{\lineheight{1.25}\smash{\begin{tabular}[t]{l}0\end{tabular}}}}%
    \put(0.59718371,0.94106325){\color[rgb]{0,0,0}\makebox(0,0)[lt]{\lineheight{1.25}\smash{\begin{tabular}[t]{l}2\end{tabular}}}}%
    \put(0.64915396,1.00600774){\color[rgb]{0,0,0}\makebox(0,0)[lt]{\lineheight{1.25}\smash{\begin{tabular}[t]{l}4\end{tabular}}}}%
    \put(0.69745757,1.06840584){\color[rgb]{0,0,0}\makebox(0,0)[lt]{\lineheight{1.25}\smash{\begin{tabular}[t]{l}6\end{tabular}}}}%
    \put(0.74281401,1.12508357){\color[rgb]{0,0,0}\makebox(0,0)[lt]{\lineheight{1.25}\smash{\begin{tabular}[t]{l}8\end{tabular}}}}%
    \put(0.78280188,1.17968439){\color[rgb]{0,0,0}\makebox(0,0)[lt]{\lineheight{1.25}\smash{\begin{tabular}[t]{l}10\end{tabular}}}}%
    \put(0.02243456,1.3030036){\color[rgb]{0,0,0}\rotatebox{90}{\makebox(0,0)[t]{\lineheight{1.25}\smash{\begin{tabular}[t]{c}$\sigma$\end{tabular}}}}}%
    \put(0.1234414,1.12014245){\color[rgb]{0,0,0}\makebox(0,0)[rt]{\lineheight{1.25}\smash{\begin{tabular}[t]{r}0.0\end{tabular}}}}%
    \put(0.10806351,1.20488805){\color[rgb]{0,0,0}\makebox(0,0)[rt]{\lineheight{1.25}\smash{\begin{tabular}[t]{r}0.5\end{tabular}}}}%
    \put(0.09044857,1.29415801){\color[rgb]{0,0,0}\makebox(0,0)[rt]{\lineheight{1.25}\smash{\begin{tabular}[t]{r}1.0\end{tabular}}}}%
    \put(0.07287714,1.39260112){\color[rgb]{0,0,0}\makebox(0,0)[rt]{\lineheight{1.25}\smash{\begin{tabular}[t]{r}1.5\end{tabular}}}}%
    \put(0.0511959,1.50545213){\color[rgb]{0,0,0}\makebox(0,0)[rt]{\lineheight{1.25}\smash{\begin{tabular}[t]{r}2.0\end{tabular}}}}%
    \put(0.2871099,0.95260262){\color[rgb]{0,0,0}\makebox(0,0)[rt]{\lineheight{1.25}\smash{\begin{tabular}[t]{r}$\lambda_6$\end{tabular}}}}%
    \put(0.50047055,0.87144285){\color[rgb]{0,0,0}\makebox(0,0)[rt]{\lineheight{1.25}\smash{\begin{tabular}[t]{r}0\end{tabular}}}}%
    \put(0.42238257,0.91791106){\color[rgb]{0,0,0}\makebox(0,0)[rt]{\lineheight{1.25}\smash{\begin{tabular}[t]{r}2\end{tabular}}}}%
    \put(0.34781105,0.96074999){\color[rgb]{0,0,0}\makebox(0,0)[rt]{\lineheight{1.25}\smash{\begin{tabular}[t]{r}4\end{tabular}}}}%
    \put(0.27840801,1.00086986){\color[rgb]{0,0,0}\makebox(0,0)[rt]{\lineheight{1.25}\smash{\begin{tabular}[t]{r}6\end{tabular}}}}%
    \put(0.21061276,1.04192316){\color[rgb]{0,0,0}\makebox(0,0)[rt]{\lineheight{1.25}\smash{\begin{tabular}[t]{r}8\end{tabular}}}}%
    \put(0.14466988,1.08024823){\color[rgb]{0,0,0}\makebox(0,0)[rt]{\lineheight{1.25}\smash{\begin{tabular}[t]{r}10\end{tabular}}}}%
    \put(0.20841922,1.59476956){\color[rgb]{0,0,0}\makebox(0,0)[lt]{\lineheight{1.25}\smash{\begin{tabular}[t]{l}(b)\end{tabular}}}}%
    \put(0.70543851,0.13474806){\color[rgb]{0,0,0}\makebox(0,0)[lt]{\lineheight{1.25}\smash{\begin{tabular}[t]{l}$p$ (a.u.)\end{tabular}}}}%
    \put(0.53988494,0.00132259){\color[rgb]{0,0,0}\makebox(0,0)[lt]{\lineheight{1.25}\smash{\begin{tabular}[t]{l}0\end{tabular}}}}%
    \put(0.5971837,0.0734823){\color[rgb]{0,0,0}\makebox(0,0)[lt]{\lineheight{1.25}\smash{\begin{tabular}[t]{l}2\end{tabular}}}}%
    \put(0.64915395,0.13842674){\color[rgb]{0,0,0}\makebox(0,0)[lt]{\lineheight{1.25}\smash{\begin{tabular}[t]{l}4\end{tabular}}}}%
    \put(0.69745756,0.2008249){\color[rgb]{0,0,0}\makebox(0,0)[lt]{\lineheight{1.25}\smash{\begin{tabular}[t]{l}6\end{tabular}}}}%
    \put(0.742814,0.25750263){\color[rgb]{0,0,0}\makebox(0,0)[lt]{\lineheight{1.25}\smash{\begin{tabular}[t]{l}8\end{tabular}}}}%
    \put(0.78280187,0.31210345){\color[rgb]{0,0,0}\makebox(0,0)[lt]{\lineheight{1.25}\smash{\begin{tabular}[t]{l}10\end{tabular}}}}%
    \put(0.02243455,0.43542275){\color[rgb]{0,0,0}\rotatebox{90}{\makebox(0,0)[t]{\lineheight{1.25}\smash{\begin{tabular}[t]{c}$\sigma$\end{tabular}}}}}%
    \put(0.12344139,0.25256151){\color[rgb]{0,0,0}\makebox(0,0)[rt]{\lineheight{1.25}\smash{\begin{tabular}[t]{r}0.0\end{tabular}}}}%
    \put(0.1080635,0.33730729){\color[rgb]{0,0,0}\makebox(0,0)[rt]{\lineheight{1.25}\smash{\begin{tabular}[t]{r}0.5\end{tabular}}}}%
    \put(0.09044856,0.42657694){\color[rgb]{0,0,0}\makebox(0,0)[rt]{\lineheight{1.25}\smash{\begin{tabular}[t]{r}1.0\end{tabular}}}}%
    \put(0.07287713,0.52502018){\color[rgb]{0,0,0}\makebox(0,0)[rt]{\lineheight{1.25}\smash{\begin{tabular}[t]{r}1.5\end{tabular}}}}%
    \put(0.05119589,0.63787125){\color[rgb]{0,0,0}\makebox(0,0)[rt]{\lineheight{1.25}\smash{\begin{tabular}[t]{r}2.0\end{tabular}}}}%
    \put(0.28710991,0.08502174){\color[rgb]{0,0,0}\makebox(0,0)[rt]{\lineheight{1.25}\smash{\begin{tabular}[t]{r}$\eta_2-\eta_1$\end{tabular}}}}%
    \put(0.50264735,0.00238077){\color[rgb]{0,0,0}\makebox(0,0)[rt]{\lineheight{1.25}\smash{\begin{tabular}[t]{r}$-$0.2\end{tabular}}}}%
    \put(0.40899223,0.05954949){\color[rgb]{0,0,0}\makebox(0,0)[rt]{\lineheight{1.25}\smash{\begin{tabular}[t]{r}$-$0.1\end{tabular}}}}%
    \put(0.31510361,0.11447968){\color[rgb]{0,0,0}\makebox(0,0)[rt]{\lineheight{1.25}\smash{\begin{tabular}[t]{r}0.0\end{tabular}}}}%
    \put(0.23152367,0.16530031){\color[rgb]{0,0,0}\makebox(0,0)[rt]{\lineheight{1.25}\smash{\begin{tabular}[t]{r}0.1\end{tabular}}}}%
    \put(0.14754184,0.21266717){\color[rgb]{0,0,0}\makebox(0,0)[rt]{\lineheight{1.25}\smash{\begin{tabular}[t]{r}0.2\end{tabular}}}}%
    \put(0.20841923,0.72718859){\color[rgb]{0,0,0}\makebox(0,0)[lt]{\lineheight{1.25}\smash{\begin{tabular}[t]{l}(c)\end{tabular}}}}%
    \put(0.96190028,1.88371951){\color[rgb]{0,0,0}\makebox(0,0)[lt]{\lineheight{1.25}\smash{\begin{tabular}[t]{l}0.0\end{tabular}}}}%
    \put(0.96190028,2.02393549){\color[rgb]{0,0,0}\makebox(0,0)[lt]{\lineheight{1.25}\smash{\begin{tabular}[t]{l}0.5\end{tabular}}}}%
    \put(0.96128743,2.16415145){\color[rgb]{0,0,0}\makebox(0,0)[lt]{\lineheight{1.25}\smash{\begin{tabular}[t]{l}1.0\end{tabular}}}}%
    \put(0.96128743,2.30436741){\color[rgb]{0,0,0}\makebox(0,0)[lt]{\lineheight{1.25}\smash{\begin{tabular}[t]{l}1.5\end{tabular}}}}%
    \put(0.96204444,2.44458336){\color[rgb]{0,0,0}\makebox(0,0)[lt]{\lineheight{1.25}\smash{\begin{tabular}[t]{l}2.0\end{tabular}}}}%
  \end{picture}%
\endgroup%

%% file: figs/gammaSigma.pdf_tex
%% Creator: Inkscape inkscape 0.92.4, www.inkscape.org
%% PDF/EPS/PS + LaTeX output extension by Johan Engelen, 2010
%% Accompanies image file 'gammaSigma.pdf' (pdf, eps, ps)
%%
%% To include the image in your LaTeX document, write
%%   \input{<filename>.pdf_tex}
%%  instead of
%%   \includegraphics{<filename>.pdf}
%% To scale the image, write
%%   \def\svgwidth{<desired width>}
%%   \input{<filename>.pdf_tex}
%%  instead of
%%   \includegraphics[width=<desired width>]{<filename>.pdf}
%%
%% Images with a different path to the parent latex file can
%% be accessed with the `import' package (which may need to be
%% installed) using
%%   \usepackage{import}
%% in the preamble, and then including the image with
%%   \import{<path to file>}{<filename>.pdf_tex}
%% Alternatively, one can specify
%%   \graphicspath{{<path to file>/}}
%% 
%% For more information, please see info/svg-inkscape on CTAN:
%%   http://tug.ctan.org/tex-archive/info/svg-inkscape
%%
\begingroup%
  \makeatletter%
  \providecommand\color[2][]{%
    \errmessage{(Inkscape) Color is used for the text in Inkscape, but the package 'color.sty' is not loaded}%
    \renewcommand\color[2][]{}%
  }%
  \providecommand\transparent[1]{%
    \errmessage{(Inkscape) Transparency is used (non-zero) for the text in Inkscape, but the package 'transparent.sty' is not loaded}%
    \renewcommand\transparent[1]{}%
  }%
  \providecommand\rotatebox[2]{#2}%
  \newcommand*\fsize{\dimexpr\f@size pt\relax}%
  \newcommand*\lineheight[1]{\fontsize{\fsize}{#1\fsize}\selectfont}%
  \ifx\svgwidth\undefined%
    \setlength{\unitlength}{1270bp}%
    \ifx\svgscale\undefined%
      \relax%
    \else%
      \setlength{\unitlength}{\unitlength * \real{\svgscale}}%
    \fi%
  \else%
    \setlength{\unitlength}{\svgwidth}%
  \fi%
  \global\let\svgwidth\undefined%
  \global\let\svgscale\undefined%
  \makeatother%
  \begin{picture}(1,0.47244094)%
    \lineheight{1}%
    \setlength\tabcolsep{0pt}%
    \put(0,0){\includegraphics[width=\unitlength,page=1]{gammaSigma.pdf}}%
    \put(0.54796784,0.23332063){\color[rgb]{0,0,0}\makebox(0,0)[rt]{\lineheight{1.25}\smash{\begin{tabular}[t]{r}\small $\Gamma_m$\end{tabular}}}}%
    \put(0.96678532,0.09573667){\color[rgb]{0,0,0}\makebox(0,0)[lt]{\lineheight{1.25}\smash{\begin{tabular}[t]{l}\small $\alpha$\end{tabular}}}}%
    \put(0.97096633,0.17146335){\color[rgb]{0,0,0}\makebox(0,0)[lt]{\lineheight{1.25}\smash{\begin{tabular}[t]{l}\small 1\end{tabular}}}}%
    \put(0.89771921,0.00907929){\color[rgb]{0,0,0}\makebox(0,0)[t]{\lineheight{1.25}\smash{\begin{tabular}[t]{c}\small 0\end{tabular}}}}%
    \put(0.72199981,0.01735604){\color[rgb]{0,0,0}\makebox(0,0)[t]{\lineheight{1.25}\smash{\begin{tabular}[t]{c}\small $\sigma^*$\end{tabular}}}}%
    \put(0.92089241,0.03542413){\color[rgb]{0,0,0}\makebox(0,0)[lt]{\lineheight{1.25}\smash{\begin{tabular}[t]{l}\small 0\end{tabular}}}}%
    \put(0.59425889,0.0584018){\color[rgb]{0,0,0}\makebox(0,0)[t]{\lineheight{1.25}\smash{\begin{tabular}[t]{c}\small 1\end{tabular}}}}%
    \put(0.57499792,0.08430846){\color[rgb]{0,0,0}\makebox(0,0)[rt]{\lineheight{1.25}\smash{\begin{tabular}[t]{r}\small 0\end{tabular}}}}%
    \put(0.55200573,0.39683351){\color[rgb]{0,0,0}\makebox(0,0)[rt]{\lineheight{1.25}\smash{\begin{tabular}[t]{r}\small 1\end{tabular}}}}%
    \put(0.55014664,0.44370079){\color[rgb]{0,0,0}\makebox(0,0)[lt]{\lineheight{1.25}\smash{\begin{tabular}[t]{l}\small (b)\end{tabular}}}}%
    \put(0,0){\includegraphics[width=\unitlength,page=2]{gammaSigma.pdf}}%
    \put(0.45234342,0.09455556){\color[rgb]{0,0,0}\makebox(0,0)[lt]{\lineheight{1.25}\smash{\begin{tabular}[t]{l}\small $\alpha$\end{tabular}}}}%
    \put(0.45534332,0.17028224){\color[rgb]{0,0,0}\makebox(0,0)[lt]{\lineheight{1.25}\smash{\begin{tabular}[t]{l}\small 1\end{tabular}}}}%
    \put(0.379734,0.00789818){\color[rgb]{0,0,0}\makebox(0,0)[t]{\lineheight{1.25}\smash{\begin{tabular}[t]{c}\small 0\end{tabular}}}}%
    \put(0.2040146,0.01617494){\color[rgb]{0,0,0}\makebox(0,0)[t]{\lineheight{1.25}\smash{\begin{tabular}[t]{c}\small $\sigma^*$\end{tabular}}}}%
    \put(0.40645051,0.03424302){\color[rgb]{0,0,0}\makebox(0,0)[lt]{\lineheight{1.25}\smash{\begin{tabular}[t]{l}\small 0\end{tabular}}}}%
    \put(0.07627368,0.05722069){\color[rgb]{0,0,0}\makebox(0,0)[t]{\lineheight{1.25}\smash{\begin{tabular}[t]{c}\small 1\end{tabular}}}}%
    \put(0.05701271,0.08430846){\color[rgb]{0,0,0}\makebox(0,0)[rt]{\lineheight{1.25}\smash{\begin{tabular}[t]{r}\small 0\end{tabular}}}}%
    \put(0.03402052,0.39801461){\color[rgb]{0,0,0}\makebox(0,0)[rt]{\lineheight{1.25}\smash{\begin{tabular}[t]{r}\small 1\end{tabular}}}}%
    \put(0.02762043,0.23332063){\color[rgb]{0,0,0}\makebox(0,0)[rt]{\lineheight{1.25}\smash{\begin{tabular}[t]{r}\small $\Gamma_l$\end{tabular}}}}%
    \put(0.02979923,0.44370079){\color[rgb]{0,0,0}\makebox(0,0)[lt]{\lineheight{1.25}\smash{\begin{tabular}[t]{l}\small (a)\end{tabular}}}}%
    \put(-0.04982696,0.53193648){\color[rgb]{0,0,0}\makebox(0,0)[lt]{\begin{minipage}{1.20662049\unitlength}\raggedright \end{minipage}}}%
  \end{picture}%
\endgroup%